\documentclass{JHEP3}

\usepackage{graphicx}
\usepackage{amssymb}
\usepackage[mathscr]{eucal}
\usepackage{amsmath}
\usepackage{cite}
\usepackage{afterpage}

\newcommand{\lta}{\lesssim}

\newcommand{\mpl}{m_{_\mathrm{Pl}}}

\newcommand{\ie}{\textsl{i.e.}~}

\newcommand{\gsimm}{\raise.3ex\hbox{$>$\kern-.75em\lower1ex\hbox{$\sim$}}}
\newcommand{\lsimm}{\raise.3ex\hbox{$<$\kern-.75em\lower1ex\hbox{$\sim$}}}

\title{Decoupling Dark Energy from Matter}

\author{Philippe Brax \\
  Institut de Physique Th\'eorique, CEA, IPhT, CNRS, URA 2306,
  F-91191Gif/Yvette Cedex, France and Institut d'Astrophysique
  de Paris, UMR 7095-CNRS, Universit\'e Pierre et Marie Curie, 98bis
  boulevard Arago, 75014 Paris, France \\ E-mail:
  \email{philippe.brax@cea.fr}}

\author{Carsten van de Bruck \\ Department of Applied Mathematics,
  University of Sheffield Hounsfield Road, Sheffield S3 7RH, United
  Kingdom \\ E-mail: \email{c.vandebruck@sheffield.ac.uk}}

\author{J\'er\^ome Martin \\ Institut d'Astrophysique de Paris, UMR
  7095-CNRS, Universit\'e Pierre et Marie Curie, 98bis boulevard
  Arago, 75014 Paris, France \\ E-mail: \email{jmartin@iap.fr}}

\author{Anne-Christine Davis \\ Department of Applied Mathematics and
  Theoretical Physics, Centre for Mathematical Sciences, Cambridge CB3
  0WA, United Kingdom \\ E-mail: \email{a.c.davis@damtp.cam.ac.uk}}

\date{today}

\abstract{We examine the embedding of dark energy in high energy
  models based upon supergravity and extend the usual phenomenological
  setting comprising an observable sector and a hidden supersymmetry
  breaking sector by including a third sector leading to the
  acceleration of the expansion of the universe. We find that
  gravitational constraints on the non-existence of a fifth force
  naturally imply that the dark energy sector must possess an
  approximate shift symmetry. When exact, the shift symmetry provides
  an example of a dark energy sector with a runaway potential and a
  nearly massless dark energy field whose coupling to matter is very
  weak, contrary to the usual lore that dark energy fields must couple
  strongly to matter and lead to gravitational
  inconsistencies. Moreover, the shape of the potential is stable
  under one-loop radiative corrections. When the shift symmetry is
  slightly broken by higher order terms in the K\"ahler potential, the
  coupling to matter remains small. However, the cosmological dynamics
  are largely affected by the shift symmetry breaking operators
  leading to the appearance of a minimum of the scalar potential such
  that dark energy behaves like an effective cosmological constant
  from very early on in the history of the universe.}

\begin{document}

\section{Introduction}
\label{sec:introduction}

Cosmological
observations~\cite{Perlmutter:1998np,Riess:1998cb,2008arXiv0803.0593N,
  2008arXiv0803.0586D,2008arXiv0803.0547K,Scranton:2003in,Tegmark:2003ud,
  Fosalba:2003ge,Solevi:2004tk,Fuzfa:2006ps} show evidence for an
accelerated expansion, usually attributed to a new form of energy,
dubbed dark energy. The simplest model for dark energy is the
cosmological constant. From the point of view of particle physics, the
cosmological constant is interpreted as the energy of the vacuum of
the universe, which must be, according to observations, 120 orders of
magnitude smaller than its "natural" value, the Planck energy
scale. So far, the existence of a pure cosmological constant has been
the most economical way of interpreting the observational data. Yet
such a tiny cosmological constant is drastically at odds with particle
physics~\cite{Weinberg:1988cp} and therefore calls for a deeper
explanation. Of course it could well be that the acceleration of the
expansion of the universe is not due to dark energy but to a large
scale modification of gravity. This possibility is under intense
scrutiny (see e.g. \cite{Dvali:2000hr,Rubakov:2008nh}). On the other
hand, if correct, the observation of a tiny vacuum energy is all the
more puzzling as the physics of phenomena at a such a low energy scale
(below the electron-Volt) is well known and has been tested in
laboratory experiments for at least 150 years. In particular, it would
seem reasonable to treat dark energy using methods which have been so
successful in describing particle physics from the atomic scale to the
weak scale. This is realised in a large class of models where dark
energy is attributed to a slowly rolling scalar field whose potential
is of the runaway
type~\cite{Wetterich:1987fm,Ratra:1987rm,Ferreira:1997hj,Liddle:1998xm}. The
field value now is of the order of the Planck mass. Within the realm
of high energy physics, this has prompted the use of supergravity
where large field values can be
handled~\cite{Brax:1999gp,Brax:1999yv,Choi:1999xn,Copeland:2000vh}.

\par

In supergravity models of particle physics, two sectors are envisaged
generically~\cite{Nilles:1983ge}. The so-called hidden sector breaks
supersymmetry leading to a splitting of masses between the
super-partners in the observable sector. In this setting the
observable sector can be taken to be the Minimal Supersymmetric
Standard Model (MSSM) whose phenomenology has been thoroughly studied
and may be discovered at the Large Hadron Collider (LHC). Moreover
supergravity models may play the role of low energy theory for a
putative unified theory like string theory~\cite{zwiebach}. Dark
energy must be included in this setting and is required to belong to a
separate sector. This is to prevent the direct couplings between dark
energy and baryons, which would lead to large deviations in tests of
Newton's
law~\cite{Will:2005va,Fischbach:1999bc,Bertotti:2003rm,Adelberger:2004ct}
and the large discrepancy between the supersymmetry breaking scale and
the vacuum energy scale.

\par

One of the first problems in this approach is the need to have a good
understanding of the theory close to its ultra-violet cut-off. Indeed,
operators suppressed by the Planck scale may affect the dark energy
predictions when the dark energy field reaches its present
value. Another issue concerns the role of radiative
corrections. Indeed the dark energy potential must be extremely flat
and radiative corrections may lift the potential altogether. The
flatness problem is particularly acute when dark energy couples to
matter and the radiative corrections induce large field-dependent
corrections. The non-field dependent corrections are also potent
although they can be absorbed in the unknown cancellation of the bare
cosmological constant. Another guise of the same problem is the
potential presence of large deviations from Newton's law when the dark
energy field is sufficiently light to mediate a fifth force between
massive bodies. Two mechanisms can be envisaged in this case. Either a
conspiracy leads to a small coupling of dark energy to matter at the
present field value, the Damour-Polyakov effect~\cite{Damour:1994zq},
or the chameleon effect where a density dependence of the dark energy
mass leads to a screening of the fifth
force~\cite{Khoury:2003rn,Brax:2004qh,Mota:2006ed}. In the following
we will address these issues in models where dark energy, the
supersymmetry breaking sector and the MSSM interact only
gravitationally. In these models, the coupling of dark energy to
gravity is generically too large. We advocate that the use of a shift
symmetry akin to the one solving the inflationary $\eta$ problem
alleviates this problem~\cite{Kawasaki:2000yn}. In inflationary
models, the shift symmetry is often interpreted either as as axionic
periodicity or as the result of the translational invariance of brane
systems in extra-dimensional models~\cite{Burgess:2007pz}. We will not
try to find the origin of the shift symmetry in the dark energy
sector, leaving it for future work.  We also study the effect of shift
symmetry breaking operators and the resulting cosmological
evolution. We find that models either lead to dark energy with a tiny
coupling to matter thanks to the shift symmetry or the shift symmetry
is broken and dark energy effectively behaves like a cosmological
constant.

\par

In the next section, we present the supergravity models of dark
energy. In section 3, we consider a dark energy sector which exhibits
a shift symmetry and analyse its effects on supersymmetry breaking,
its coupling to the MSSM and study the deviations from Newton's
law. We find that the presence of a shift symmetry is strongly
motivated and gives a direct link between the dark energy potential
and the superpotential of the model. In section 4, we break this shift
symmetry by non-renormalisable operators in the K\"ahler potential and
find that the resulting cosmology is akin to a pure cosmological
constant. In section 5, we discuss briefly a different ansatz for the
superpotential. Our findings are summarised in the last section.

\section{Dark Energy and Shift Symmetry}
\label{sec:deshift}

We are interested in the coupling between matter, either baryonic or
dark, and dark energy in the context of particle physics models
connected to high energy physics. More precisely, we shall be
concerned with supersymmetric models beyond the standard model of
particle physics. These models provide a convenient framework within
which one may study some aspects of string theory (the low energy
effective action) or describe possible observable consequences at the
LHC.

\par

Supersymmetry has not yet been observed hence it must be broken at a
scale which may be a few TeV in order to address the hierarchy problem
between the Planck scale and the electro-weak scale. Supersymmetry is
usually assumed to be broken in a hidden sector. In the following, we
will concentrate on the possibility that the breaking of supersymmetry
is transmitted to the standard model via gravitational
interactions. Other mechanisms such as gauge mediation are popular and
require a special treatment which is left for future work. Therefore,
the usual structure of the standard model, the MSSM, is as
follows. There are two sectors, the observable sector and the hidden
sector and the assumption that the two sectors interact only
gravitationally is expressed by~\cite{Nilles:1983ge}
\begin{eqnarray}
K &=&  K_{_{\rm MSSM}}\left(\phi^a,\phi^{\bar a\dagger }\right)+
K_{\rm h} \left(z,z^{\dagger}\right)\, ,\quad
W = W_{_{\rm MSSM}}(\phi^a)+ W_{\rm h} (z)\, ,
\end{eqnarray}
where we restrict ourselves to a single field $z$ in the hidden
sector. Explicitly, the MSSM sector is defined by
\begin{equation}
K_{_{\rm MSSM}}=\phi ^a\phi^{\bar a}{}^{\dagger}\, ,\quad
W_{_{\rm MSSM}}= \frac{1}{3} \lambda_{abc} \phi^a\phi^b\phi^c
+\frac{1}{2} \mu_{ab} \phi^a\phi^b\, ,
\end{equation}
where $\lambda_{abc}$ are the Yukawa couplings and $\mu_{ab}$ has a
single non-zero entry  for the two Higgs bosons $\mu_{H_{\rm
    u}H_{\rm d}}\equiv \mu$. In this framework, the breaking of
supersymmetry is parameterised by the $F$-term
\begin{equation}
F_z= {\rm e}^{\kappa_4^2 K_{\rm h}/2} D_z W_{\rm h}
\end{equation}
at the minimum of the hidden sector potential, $\langle z\rangle
=z_0$. In the above equation, we have used the definition $D_z W =
\partial_zW + \kappa_4^2 K_z W$, where $\kappa_4^2= 1/M_{\rm Pl}^2= 8\pi
G_{\rm N}$ and $K_z = \partial_z K$.
In fact, the $F_z$ term is defined in terms of the covariant
derivative of the total superpotential. At high energy, the
fields of the observable sector vanish and $W\simeq W_{\rm h}$. In the
absence of dark energy, the bare cosmological constant is set to zero at
tree level by requiring the cancellation at the minimum
\begin{equation}
\vert F_z \vert_{z_0} = \sqrt 3 m_{3/2}\vert _{z_0}\, ,
\end{equation}
where the gravitino mass is defined to be
\begin{equation}
m_{3/2}\equiv \kappa_4 W_{\rm h}{\rm e}^{\kappa_4^2 K_{\rm h}/2}\, .
\end{equation}
Once supersymmetry is broken, the superpartners of the standard model
fields acquire a mass. In the standard scenario, the running of the
masses in the Higgs sector is such that the electroweak symmetry is
broken at a low scale compared to the large scale, say the Grand
Unified Theory (GUT) scale where supersymmetry breaking takes
place. At the electroweak scale, the two Higgs fields pick up a vacuum
expectation value (vev, which we denote by $v_{\rm u}$ and $v_{\rm
  d}$) such that the masses of the fermions become
\begin{equation}
m_{\rm u,d}= \lambda {\rm e}^{\kappa_4^2 K/2}v_{\rm u,d}\, ,
\end{equation}
where $\lambda$ is the appropriate Yukawa coupling for a particle
of type $u$ or $d$.

\par

Then, the main issue is to understand how dark energy can be
implemented into the above framework. Dark energy cannot belong to the
observable sector as this would lead to a strong fifth force signal
unless the coupling constants are artificially tuned to be small. We
discard this possibility. Dark matter and dark energy could belong to
the same sector. Couplings between dark matter and dark energy have
been studied in the past (see
e.g.~\cite{Wetterich:1994bg,Amendola:1999er,Farrar:2003uw}).  However,
in the MSSM, dark matter belongs to the observable sector. As a
result, dark matter and dark energy will have only gravitational
interactions. Finally, dark energy could belong to the supersymmetry
breaking sector. In this work, we will assume that dark energy and the
breaking of supersymmetry occur in separate sectors. This is motivated
by the fact that supersymmetry breaking happens at a very large scale
compared to the dark energy scale.

\par

Therefore, one has to assume that there is a separate dark energy
sector characterised by its own K\"ahler and superpotentials,
$K_{_{\rm DE}}$ and $W_{_{\rm DE}}$. In the absence of the hidden
sector and of the MSSM, the dark energy sector would be governed by a
scalar potential
\begin{equation}
\label{eq:potde}
V_{_{\rm DE}} = {\rm e}^{\kappa_4^2 K_{_{\rm DE}}} \left(
\vert D_Q W_{_{\rm DE}}\vert^2 - 3 \kappa_4^2
\vert W_{_{\rm DE}} \vert^2~\right),
\end{equation}
which is a typical quintessence potential of the runaway type. However,
as already mentioned, the dark energy sector cannot be considered as
completely isolated since it always interacts gravitationally with the
rest of the world.

\par

All in all, our starting point will be the following separated
K\"ahler- and
superpotentials~\cite{Brax:2006dc,Brax:2006kg,Brax:2006np}
\begin{eqnarray}
K &=&  K_{_{\rm MSSM}}\left(\phi^a,\phi^{\bar a\dagger }\right)+
K_{\rm h} \left(z,z^{\dagger}\right)+K_{_{\rm DE}}
\left(Q,Q^{\dagger}\right)\, ,\\
W &=& W_{_{\rm MSSM}}(\phi^a)+ W_{\rm h}(z)+ W_{_{\rm DE}}(Q)\, ,
\end{eqnarray}
where we restrict ourselves to a single field $Q$ in the dark energy
sector for convenience.

\par

Let us now investigate what follows from the above equations. At high
energy, the hidden sector field (still assumed to be stabilised) picks
up a vev which is perturbed by the coupling to the dark energy
sector. The perturbation is small enough (the dark energy scale is
tiny compared to the supersymmetry breaking scale) to guarantee the
existence of the perturbed minimum
\begin{equation}
\langle z \rangle = z_0(Q,Q^\dagger) \, .
\end{equation}
Roughly speaking, this has two types of consequences. Firstly, the
shape of the potential controlling the evolution of dark energy which,
if the dark sector were isolated, would be given by
Eq.~(\ref{eq:potde}) can be changed by supersymmetry breaking
terms. Secondly, the gravitational physics at low energy is
affected. The interaction between the hidden and dark sectors implies
that all the soft breaking terms acquire a $Q$-dependent form. After
renormalisation down to lower scales, the Higgs potential becomes
$Q$-dependent, implying that both $v_{\rm u,d}$ become $Q$-dependent
too. This has a drastic effect on the gravitational physics at low
energy. Indeed the masses of the standard model fermions becomes
$Q$-dependent
\begin{equation}
m_{\rm u,d}(Q)=\lambda {\rm e}^{\kappa_4^2 K(Q,Q^\dagger)/2} v_{\rm u,d}(Q)~,
\end{equation}
where we have explicitly assumed that the Yukawa couplings are dark
energy independent. Another important effect is the $Q$-dependence of
the Quantum Chromo-Dynamics (QCD) scale $\Lambda$. Indeed the
superpartners of the gauge bosons, the gauginos, acquire a mass only
if the gauge coupling function $f$ such that $\Re f=1/g^2$ depends on
$z$, see Eq.~(2.24) in Ref.~\cite{Brax:2006dc}.

\par

Let us combine all these results. If the dark energy potential is
modified such that the mass of the quintessence field is now larger
than $10^{-3}\mbox{eV}$, then the problem with gravitational physics
are evaded since the range of the corresponding force is too small to
lead to effects that can be seen experimentally. However, in this
case, the potential is not of the runaway form and the model is
generally not interesting from the cosmological point of view. On the
contrary, if the hidden sector is such that the dark energy potential
preserves its shape, and, therefore, such that the model provides an
interesting alternative to the cosmological constant, one encounters
new problems. Indeed, in most runaway dark energy models, the dark
energy field has a large value now, of the order of the Planck
mass. Moreover the mass of the dark energy field is tiny and of the
order of the Hubble rate now $H_0\sim 10^{-43} \mbox{GeV}\ll
10^{-3}\mbox{eV}$. Such a low mass for a particle coupled to matter
leads to gravitational problems and the existence of a detectable
fifth force. This can be prevented if the coupling to gravity is small
as imposed by the Cassini experiment~\cite{Bertotti:2003rm}. Let us
analyse the magnitude of the gravitational coupling of dark energy to
matter. The gravitational coupling of a particle is simply given
by~\cite{Damour:1992we}
\begin{equation}
\kappa_4\alpha=\frac{{\rm d}\ln m }{{\rm d}Q_{\rm n}}\, ,
\end{equation}
where $Q_{\rm n}$ is the normalised dark energy field such that
$\partial_{Q}\partial_{\bar Q} K_{_{\rm DE}}(\partial Q)^2 = (\partial
Q_{\rm n})^2/2$ and we have chosen $Q=Q^{\dagger}$ to be real.

\par

The previous considerations would imply many low energy gravitational
effects. In particular, one expects that the weak equivalence
principle is violated at the microscopic level. Indeed, the masses of
the $u$ and $d$ fermions have a different coupling $\alpha_{\rm u}\ne
\alpha_{\rm d}$ implying that, in a gedanken experiment, particles of
type $u$ and $d$ would fall at a different speed in a constant
gravitational field. Of course, gravitational experiments are not
carried out on microscopic particles but on macroscopic objects
composed of many atoms. The mass of a particular atom can be
decomposed as~\cite{Damour:1992we}
\begin{equation}
m_{_{\rm ATOM}}\simeq M \Lambda _{_{\rm QCD}}+ \sigma' \left(N+Z\right)
+ \delta' \left(N-Z\right) + a_3
\alpha_{_{\rm QED}} E_{\rm A} \Lambda_{_{\rm QCD}} \, ,
\end{equation}
where $\Lambda _{_{\rm QCD}}\simeq 180\mbox{MeV}$ is the QCD scale,
$N$ the number of neutrons and $Z$ the number of protons. The quantity
$M$ can be written $M= (N+Z) + E_{_{\rm QCD}}/\Lambda$ where $
E_{_{\rm QCD}}$ is the strong interaction contribution to the nucleus
binding energy. The number $E_{\rm A}$ is given by $E_{\rm A}=
Z(Z-1)/(N+Z)^{1/3}$ and the quantity $a_3\alpha _{_{\rm QED}}\Lambda
_{_{\rm QCD}}E_A$ represents the Coulomb interaction of the nucleus
where $a_3\alpha_{_{\rm QED}}\simeq 0.77\times 10^{-3}$. Finally the
coefficients $\delta'$ and $\sigma'$ depend on the constituent masses
and can be expressed as
\begin{eqnarray}
\label{eq:mess1}
\sigma'&=& \frac{1}{2}(m_{\rm u}+m_{\rm d})(b_{\rm u}+b_{\rm d})
+\frac{\alpha_{_{\rm
QED}}}{2}(C_{\rm n}+C_{\rm p}) +\frac{1}{2} m_{\rm e}\, , \\
\label{eq:mess2}
\delta'&=&\frac{1}{2}(m_{\rm u}-m_{\rm d})(b_{\rm u}-b_{\rm d})
+\frac{\alpha_{_{\rm QED}}}{2} (C_{\rm n}-C_{\rm p}) -\frac{1}{2} m_{\rm
e}\, ,
\end{eqnarray}
where $m_{\rm u}\sim 5~{\rm MeV}$, $m_{\rm d} \sim {10}~{\rm MeV}$. The
constants appearing in Eqs.~(\ref{eq:mess1}) and~(\ref{eq:mess2}) are
given by: $b_{\rm u}+b_{\rm d}\simeq 6$, $b_{\rm u}-b_{\rm d}\sim 0.5$,
$C_{\rm p} \alpha_{_{\rm QED}} \simeq 0.63~{\rm MeV}$, $C_{\rm
n}\alpha_{_{\rm QED}}\sim -0.13~{\rm MeV}$. This implies that
$\sigma'/\Lambda_{_{\rm QCD}}\simeq 3.8 \times 10^{-2}$ and
$\delta'/\Lambda_{_{\rm QCD}} \simeq 4.2\times 10^{-4}$.

\par

The fact that $m_{\rm u}$ and $m_{\rm d}$ are dark energy-dependent
quantities imply that the coefficients $\alpha '$ and $\delta '$, and
hence $m_{_{\rm ATOM}}$, are now $Q$-dependent quantities. However,
this is not the only source of $Q$-dependence. Indeed, the low energy
gauge couplings are given by
\begin{equation}
\frac{1}{\alpha_i(m)}= 4\pi f_i -\frac{b_i}{2\pi} \ln \left(\frac{m_{\rm
GUT}}{m}\right)\, ,
\end{equation}
where $i=1, \cdots, 3$ for $U(1)_{\rm Y}$, $SU(2)_{\rm L}$ and $SU(3)$
respectively with $b_i = (-33/5,-1,3)$. The quantity $f_i$ is the
gauge coupling function already discussed before. This implies that
the QCD scale is related to the gauge coupling function as
\begin{equation}
\Lambda_{_{\rm QCD}}= m_{_{\rm GUT}} {\rm e}^{-8\pi^2f_3/b_3},
\end{equation}
where we have assumed gauge coupling unification at $m_{_{\rm
GUT}}$. Since $f_3$ is a function of $Q$, $\Lambda_{_{\rm QCD}}$ is also
a $Q$-dependent quantity. The same reasoning is true for $\alpha
_{_{\rm QED}}$ since
\begin{equation}
\alpha_{_{\rm QED}}= \frac{\alpha_2^2}{\alpha_1+\alpha_2}\, ,
\end{equation}
$\alpha _1$ and $\alpha _2$ being related to $f_1$ and $f_2$.

\par

We are now in a position where one can estimate the typical
gravitational coupling of an atom. From the previous considerations,
one obtains
\begin{eqnarray}
\kappa_4 \alpha_{_{\rm ATOM}} &=& -\frac{8\pi^2}{b_3} \frac{\partial
f_3}{\partial Q_{\rm n}} +\frac{N+Z}{M} \frac{\partial}{\partial
Q_{\rm n}}\left(\frac{\sigma'}{\Lambda_{_{\rm QCD}}}\right) +\frac{N-Z}{M}
\frac{\partial}{\partial Q_{\rm n}}
\left(\frac{\delta'}{\Lambda_{_{\rm QCD}}}\right)
\nonumber \\
&+& a_3 \frac{E_A}{M} \frac{\partial \alpha_{_{\rm QED}}}{\partial Q_{\rm n}}~,
\end{eqnarray}
where the variations of $\sigma'$ and $\delta'$ read
\begin{eqnarray}
\frac{\partial}{\partial Q_{\rm n}}\left(\frac{\sigma'}
{\Lambda_{_{\rm QCD}}}\right)&=&
\frac{\kappa_4}{2\Lambda_{_{\rm QCD}}} (b_{\rm u}+b_{\rm d})
\left(\alpha_{\rm u} m_{\rm u} +\alpha_{\rm d}m_{\rm d}\right)
+\frac{\kappa_4}{2\Lambda_{_{\rm QCD}}} \alpha_{\rm d} m_{\rm e}
\nonumber \\ & &
+\frac{8\pi^2}{b_3}
\frac{\sigma'}{\Lambda_{_{\rm QCD}}}\frac{\partial f}{\partial Q_{\rm n}}
+ \frac{C_{\rm n}+C_{\rm p}}{2\Lambda_{_{\rm QCD}}}
\frac{\partial \alpha_{_{\rm QED}}}{\partial
Q_{\rm n}}~, \nonumber \\
\frac{\partial}{\partial Q_{\rm n}}
\left(\frac{\delta'}{\Lambda _{_{\rm QCD}}}\right)&=&
-\frac{\kappa_4}{2\Lambda _{_{\rm QCD}}} \left(b_{\rm u}-b_{\rm d}\right)
\left(\alpha_{\rm u} m_{\rm u} -\alpha_{\rm d}m_{\rm d}\right)
-\frac{\kappa_4}{2\Lambda_{_{\rm QCD}}} \alpha_{\rm d} m_{\rm e}
\nonumber \\ & &
+ \frac{8\pi^2}{b_3}
\frac{\sigma'}{\Lambda_{_{\rm QCD}}}\frac{\partial f}{\partial Q_{\rm
n}} + \frac{C_{\rm n}-C_{\rm p}}{2\Lambda_{_{\rm QCD}}}
\frac{\partial \alpha_{_{\rm QED}}}{\partial
Q_{\rm n}}~,
\end{eqnarray}
with the coefficients $\alpha_{\rm u}$ and $\alpha_{\rm d}$ given by
\begin{eqnarray}
\alpha_{\rm u} &=& \frac{\kappa_4}{2} \partial_{Q_{\rm n}} K_{_{\rm DE}} +
\frac{\kappa_4}{2} \partial_{Q_{\rm n}} K_{\rm h} +
\frac{\kappa_4}{v_{\rm u}}\frac{{\rm d} v_{\rm u}}{{\rm d} Q_{\rm n}}~,
\\
\alpha_{\rm d} &=& \frac{\kappa_4}{2} \partial_{Q_{\rm n}} K_{_{\rm DE}} +
\frac{\kappa_4}{2} \partial_{Q_{\rm n}} K_{\rm h} +
\frac{\kappa_4}{v_{\rm d}}\frac{{\rm d} v_{\rm d}}{{\rm d} Q_{\rm n}}~,
\end{eqnarray}
and similarly the fine structure constant has a variation induced by
the $Q$-dependence of the hidden sector vev $z_0\left(Q_{\rm n}\right)$
\begin{equation}
\frac{\partial \alpha_{_{\rm QED}}}{\partial Q_{\rm n}}= 4\pi\left[
\frac{\alpha_1^2\alpha_2^2-(2\alpha_1 +\alpha_2)\alpha_2^3}
{(\alpha_1+\alpha_2)^2}\right]\frac{\partial f}{\partial Q_{\rm n}}~.
\end{equation}
Although complicated, the above expressions allow us to compute
$\alpha_{_{\rm ATOM}}$ exactly for the MSSM model. In fact, the
analysis can be simplified if one notices that the leading effect in
the gravitational coupling $\alpha_{_{\rm ATOM}}$ comes from the bare
dependence on the dark energy K\"ahler potential
\begin{eqnarray}
\label{eq:alphaatom}
\alpha_{_{\rm ATOM}} &\simeq & \left[ \frac{N+Z}{M}
\frac{m_{\rm u}+m_{\rm d}}{4\Lambda_{_{\rm QCD}}}
\left(b_{\rm u}+b_{\rm d}\right) -\frac{N-Z}{M}
\frac{m_{\rm u}-m_{\rm d}}{4\Lambda_{_{\rm QCD}}}\left(b_{\rm u}-b_{\rm d}\right)
\right]\nonumber \\ & & \times \kappa_4 \partial_{Q_{\rm n}}
K_{_{\rm DE}} +\cdots
\end{eqnarray}
Despite the smallness of the quark masses compared to the QCD scale,
the prefactor of $\kappa_4 \partial_{Q_{\rm n}} K_{_{\rm DE}}$ is no
less than $10\%$. Let us now consider simple examples where the main
trend can be grasped. For a canonically normalised field, we find that
$K_{_{\rm DE}}= QQ^{\dagger}$, leading to $Q_{\rm n}= \sqrt{2} Q$ and
\begin{equation}
  \partial_{Q_{\rm n}} K_{_{\rm DE}}= Q_{\rm n}/2~.
\end{equation}
The main constraint on the presence of a fifth force comes from the
Cassini probe. The Cassini experiment leads to a bound on $\vert
\alpha_{_{\rm ATOM}}\vert \lta 10^{-3}$~\cite{Bertotti:2003rm}. As we
can see from the previous expressions, this has drastic consequences
for dark energy. Indeed, this implies that the value of $Q_{\rm n}$
now must be less than $10^{-2} \mpl$ to satisfy the Cassini bound. In
all dark energy models based on runaway potentials, the value of the
quintessence field is of the order of the Planck mass now. The above
expression shows that this leads to a strong violation of the Cassini
bound. Another interesting case is provided by the dilaton ($n=1$) or
moduli fields ($n=3$), with $K_{_{\rm DE}}= -n \ln \left[
  \kappa_4\left(Q+Q^{\dagger}\right) \right]$ leading to $Q_{\rm n}=
\sqrt{n/2}\ln \left(\kappa_4 Q\right)$ and
\begin{equation}
\partial_{Q_{\rm n}} K_{_{\rm DE}}= \sqrt {2n}~.
\end{equation}
This implies again a large violation of the Cassini bound.

\par

From the previous considerations, it is now clear that the fact that
the gravitational coupling is large has a similar origin to the so
called $\eta$ problem in supergravity inflation, where a supergravity
correction depending on the K\"ahler potential of the inflaton leads
to a large mass for the inflaton. In the dark energy context, the
gravitational coupling problem springs from the supergravity
correction to the fermion mass which appears as an exponential of the
K\"ahler potential. A solution to the $\eta$ problem is provided by a
shift symmetry implying that the K\"ahler potential of the inflaton
vanishes along the inflationary direction.  Similarly a solution to
the gravitational coupling constant problem can be obtained provided
the first derivative of the dark energy K\"ahler potential vanishes
identically along the dark energy direction. In practice this implies
that
\begin{equation}
K_{_{\rm DE}}\left(Q,Q^{\dagger}\right) = K_{_{\rm DE}}
\left(Q-Q^{\dagger}\right)~,
\end{equation}
where we expand $K_{_{\rm DE}}(x)= -x^2/2+\cdots$ in powers of
$x$. Notice that $Q\to Q+c$ with $c$ real is a shift symmetry of the
K\"ahler potential. In this case, the main contribution to the
gravitation coupling vanishes. As a result the gravitational coupling
depends only on the hidden sector dynamics and its coupling to the
dark energy sector.

\par

Let us also notice that another motivation for a shift symmetric K\"
ahler potential is the presence of large corrections to the dark
energy potential coming from the coupling to the hidden sector
\begin{equation}
\delta V_{_{\rm DE}}= m_{3/2}^2 K^{QQ^{\dagger}}_{_{\rm DE}} \partial_ Q
K_{_{\rm DE}} \partial_{Q^{\dagger}}K_{_{\rm DE}}~.
\end{equation}
For a canonical K\"ahler potential, this leads to a mass equal to the
gravitino mass for the dark energy field. Such a large mass implies
that the dark energy potential develops a minimum where the dark
energy field is stuck very early on in the history of the universe,
hence acting as an effective cosmological constant.  In the moduli
(dilaton) case, the correction to the potential is exponential with a
large prefactor resulting in a large contribution to the energy
density when the dark energy field is of the order of the Planck
scale; such a large value needs to be compensated by one further
tuning on top of the bare cosmological constant fine-tuning. When the
K\"ahler potential is shift symmetric, the correction to the dark
energy potential vanishes identically. Smaller corrections exist
though. We will address the question of their origin in the following.

\section{Implementing the Shift Symmetry}
\label{sec:implementshift}

In this section, our goal is to recompute the coefficient $\alpha
_{_{\rm ATOM}}$ in the case where the dark energy sector is shift
symmetric and to show that, in this case, the Cassini bound can be
easily satisfied. Therefore, we take
\begin{equation}
K_{_{\rm DE}}(Q)= -\frac{1}{2}\left(Q-Q^{\dagger}\right)^2\, ,\quad
W_{_{\rm DE}}(Q)=w(Q)~\, ,
\end{equation}
where $w(Q)$ is, for the moment, an arbitrary function. The next step
is to solve for the vev of the hidden sector field $\left\langle z
\right \rangle$. To be completely explicit, let us focus on the
coupling of dark energy to a Polonyi model with~\cite{Polonyi:1977pj}
\begin{equation}
K_{\rm h}\left(z,z^{\dagger}\right) = \vert
z\vert^2\, ,\quad
W_{\rm h}(z)= m^2(z+\beta)\, ,
\end{equation}
where $m^2\sim m_{3/2}\mpl $. We focus on the real direction
$Q=Q^{\dagger}$ as the imaginary direction is massive with a mass of
order $m_{3/2}$. Along this direction, the scalar potential reads
\begin{eqnarray}
  V(Q,z) &=& {\rm e}^{\kappa_4^2 \vert z\vert^2} \left[ \vert w'\vert^2 + \vert
    m^2\left(1+\kappa_4^2 \vert z\vert^2\right) + \kappa_4 z^{\dagger}
    \left(\kappa_4 m^2 \beta +
      \kappa_4 w\right) \vert^2 \nonumber \right. \\
  &-& \left. 3 \vert m^2 \kappa_4 z+ \kappa_4 m^2 \beta + \kappa_4
    w\vert^2 \right]~,
\end{eqnarray}
where the prime denotes a derivative with respect to $Q$.  We are
looking for the minimum of the scalar potential along the hidden
sector direction $z$ which is stabilised at
\begin{equation}
  \kappa_4z_0=1-\kappa_4\beta\, , \quad \kappa_4 z_0= \sqrt 3 -1\, ,
\end{equation}
in the absence of dark energy. When dark energy is present, the
minimum is perturbed and becomes $z_{\rm min}(Q)= z_0 +\delta z(Q)$ where
\begin{equation}
\label{eq:darkenergymin}
\delta z(Q)=\left(\sqrt{3} -1\right) \frac{w}{m^2}~,
\end{equation}
where we have neglected higher order terms in $w$ and $w'$ (see
section 4). This perturbation is very small due to the discrepancy
between the dark energy scale and the supersymmetry breaking
scale. The potential at the minimum becomes a sole function of $Q$ and
can be expressed as
\begin{equation}
\label{eq:darkenergypot}
V_{_{\rm DE}}(Q)= -2\sqrt 3 {\rm e}^{\kappa_4^2 \vert z_0\vert^2}
\kappa_4 m^2 w(Q)~.
\end{equation}
A simple method to obtain this equation is to remark that
$V=V(z_0)+V'(z_0)\delta z+V''(z_0)/2(\delta z)^2+\cdots \simeq V(z_0)$
since $V'(z_0)=0$ by definition and we work at first order in $\delta
z$. In order to guarantee the positivity of the dark energy potential,
one must impose $w<0$. As a result, for any negative and runaway
superpotential, we have found that there is a corresponding dark
energy model with a potential energy proportional to the
superpotential. Let us evaluate the order of magnitude of $w_{\rm
  now}$. We find
\begin{equation}
\vert w_{\rm now}\vert \sim \frac{\rho_{\rm cri}}{m_{3/2}}\, ,
\end{equation}
where $\rho_{\rm cri}$ is the present day critical energy
density. Therefore, one can check that $\delta z/z_0\simeq \rho _{\rm
  cri}/(m_{3/2}^2\mpl ^2)\ll 1$ (we have used the fact that the
unperturbed $z_0\simeq \mpl$) and this justifies the approximation
made above. In fact, the previous calculation is an estimate of the
value of the small dimensionless parameter used in order to perform
the perturbative expansion. Indeed, the small parameter is then given
by $\kappa_4w/m^2\simeq \rho_{\rm cri}/\mpl^4(\mpl/m_{3/2})^2\simeq
10^{-88}(100\mbox{GeV}/m_{3/2})^2$. Notice that the dimensionless
parameter $w'/m^2$ also appears in the calculation. It should be
considered of the same order as $\kappa_4w /m^2$ since $(\kappa_4w
/m^2)/(w'/m^2)\simeq \kappa_4Q\simeq 1$ now for a runaway potential.

\par

Let us give a simple example where the potential can be evaluated. The
gaugino condensation superpotential for $N_f$ flavours of quarks in
the fundamental representation of the $SU(N_c)$ gauge group can be
expressed as (the sign of the superpotential depends on a choice of
the phase of the meson matrix)~\cite{Binetruy:1998rz}
\begin{equation}
w(Q)= -(N_{\rm c}-N_{\rm f})\frac{\Lambda^{(3N_{\rm c}-N_{\rm f})/(N_{\rm c}-N_{\rm f})}}
{Q^{2N_{\rm f}/(N_{\rm c}-N_{\rm f})}}~,
\end{equation}
where $\Lambda$ is a strong interaction scale and $Q$ is the field
along the diagonal meson direction. In this case, the dark energy
potential is a Ratra-Peebles potential $V(Q)=M^{4+n}/Q^n$ where $n=
2N_{\rm f}/(N_{\rm c}-N_{\rm f})$ and $M^{4+n}=2\sqrt 3{\rm
  e}^{\kappa_4^2 \vert z_0\vert^2} \kappa_4 m^2 \Lambda^{(3N_{\rm
    c}-N_{\rm f})/(N_{\rm c}-N_{\rm f})}$. The dark energy scale is
completely specified by the supersymmetry breaking scale and the
strong interaction scale.

\par

It is also interesting to notice that, usually, when supergravity is
used to construct models of dark energy, one does not obtain the
Ratra-Peebles potential but the SUGRA potential~\cite{Brax:1999gp},
$V(Q)={\rm e}^{\kappa_4^2Q^2}M^{4+n}/Q^n$. Since the exponential
correction directly originates from a supergravity term of the form
${\rm e}^{\kappa_4^2K_{_{\rm DE}}}$, it is clear that, in the presence
of a shift symmetry, this term is not recovered. This is why, here,
one obtains the Ratra-Peebles potential exactly. In fact, this is
rather unfortunate since it is known that the dark energy equation of
state of the SUGRA potential, thanks to the exponential factor, is
much closer to $-1$ (more precisely $w_{_{\rm DE}}\simeq -0.86$, see
Ref.~\cite{Brax:1999gp}) and, hence, more compatible with the present
day observations, than the equation of state of the Ratra-Peebles
potential. This last one is indeed too far from $-1$ to be compatible
with the constraints on $w_{_{\rm DE}}$ unless one considers very
small values of $n$ which seems pretty contrived. Therefore, the fact
that, in the presence of a shift symmetry, one loses the exponential
correction in the dark energy potential should be considered as a
drawback. In other words, although the shift symmetry has solved the
$\eta$-problem, contrary to the case of inflation this does not lead
to desired features for dark energy. However, one can also consider
other forms for $w(Q)$ such as~\cite{Albrecht:1999rm}
\begin{equation}
w(Q)= \Lambda^3 \left[A+\left(\kappa_4Q-B\right)^\alpha\right]
{\rm e}^{-\lambda \kappa_4 Q}\, ,
\end{equation}
where $\Lambda $ is an energy scale and $A$, $B$ and $\alpha $ free
parameters. This would lead to a potential of the Albrecht-Skordis
type with interesting phenomenological properties such as a low value
for the equation of state. Of course, almost any shape for the
superpotential can be invoked as long as the resulting equation of
state is low enough. Eventually, one would like to have an intrinsic
justification for a given superpotential coming from a more
fundamental theory.


\par

As discussed above, the $Q$ dependence of the atomic masses appears
via the $Q$-dependence of the Higgs vev. In the supersymmetric
context, the electroweak symmetric breaking is radiatively induced as
the Higgs masses evolve from the GUT scale to the weak
scale~\cite{Derendinger:1983bz}. One of the Higgs masses becomes
negative, triggering the symmetry breaking. Therefore, one needs to
compute $v_{\rm u}$ and $v_{\rm d}$ in the shift symmetric case. At
the GUT scale, where SUSY is broken, the observable potential is
corrected by the soft supersymmetry breaking terms as follows
\begin{eqnarray}
  V_{_{\rm MSSM}}&=& {\rm e}^{\kappa_4^2 K} V_{\rm susy} + A_{abc}
  \left(\phi_a\phi_b\phi_c
+ \phi_a^{\dagger}\phi_b^{\dagger}\phi_c ^{\dagger}\right)+ B_{ab}
\left(\phi_a\phi_b+\phi _a^{\dagger}\phi_b^{\dagger}\right) \nonumber \\
& & + m^2_{a\bar
    b} \phi^a\phi^{\bar b}{}^{\dagger}\, ,
\end{eqnarray}
where $V_{\rm susy}$ is the potential in the absence of supersymmetry
breaking. In the shift symmetric case, the soft terms are explicitly
given by
\begin{eqnarray}
A_{abc}&=&\frac{\lambda_{abc}}{3}{\rm e}^{\kappa_4^2 \vert
z_0(Q)\vert^2}\left[\left(M_{_{\rm S}} +\kappa_4^2 w^{\dagger}\right)
\kappa_4^2\vert z_0(Q)\vert^2 +\kappa_4^2 m^2 z_0(Q)\right]\, ,
\nonumber\\
B_{ab}&=&\frac{\mu_{ab}}{2}{\rm e}^{\kappa_4^2 \vert z_0(Q)\vert^2}
\left[\left(M_{_{\rm S}}
+\kappa_4^2 w^{\dagger}\right)\left(\kappa_4^2\vert z_0(Q)\vert^2-1\right)
+\kappa_4^2 m^2 z_0(Q)\right]\, ,\nonumber \\
m^2_{a\bar b}& = & m_{3/2}^2(Q)\delta_{a\bar b}\, , \nonumber
\end{eqnarray}
where the gravitino mass can be expressed as
\begin{equation}
m_{3/2}(Q)= {\rm e}^{\kappa_4^2\vert z_0(Q)\vert^2/2}
\vert M_{_{\rm S}} + \kappa_4^2 w \vert\, ,
\end{equation}
and we have defined
\begin{equation}
M_{_{\rm S}}= \kappa_4^2 \left \langle W_{\rm h} \right\rangle  (Q)\, .
\end{equation}
A crucial point is already apparent here. The soft terms become
$Q$-dependent but only through the $Q$-dependence of $z_0$ and
$w(Q)$. In particular, in the non shift symmetric case, the soft terms
are all proportional to ${\rm e}^{\kappa _4^2K_{_{\rm DE}}}\simeq {\rm
  e}^{\kappa _4^2Q^2}$, see Eqs.~(2.22), (2.23) and (2.24) of
Ref.~\cite{Brax:2006dc}, which is responsible for the large violation
of the Cassini bound. Here, thanks to the shift symmetry, this factor
is absent.

\par

Let us now specialise the above formula for the observable potential
to the Higgs sector. One obtains
\begin{eqnarray}
  V_{\rm Higgs} &=& \left(\vert \mu\vert^2 {\rm e}^{\kappa_4^2 \vert z\vert^2} +
    m_{H_{\rm u}}^2\right) v_{\rm u}^2 + \left(\vert \mu\vert^2
      {\rm e}^{\kappa_4^2 \vert z\vert^2}
      + m_{H_{\rm d}}^2 \right) v_{\rm d}^2 \nonumber \\
    &-& 2 \mu B v_{\rm u}v_{\rm d} +\frac{1}{8}
    \left(g_1^2+g_2^2\right)\left(v_{\rm u}^2-v_{\rm d}^2\right)^2~,
\end{eqnarray}
where we have used that $m_{1\bar{1}}^2=m_{H_{\rm u}}^2$,
$m_{2\bar{2}}^2=m_{H_{\rm d}}^2$ with $m_{H_{\rm u}}=m_{H_{\rm
    d}}=m_{3/2}$ at the GUT scales and $B_{ab}=\mu B\epsilon
_{ab}$. The soft terms $A_{abc}$ are hidden in the above formula and
appear in the two loop expression for the renormalised Higgs masses,
see Eqs.~(3.14) and~(3.15) of Ref.~\cite{Brax:2006dc}. As already
mentioned, all the soft terms depend on $Q$ and, as they are
renormalised to low energy, they keep an intricate
$Q$-dependence. Then, one has to minimise the above potential in order
to find the expression of $v_{\rm u}$ and $v_{\rm d}$. But, thanks to
the shift symmetry, the $Q$-dependence of the minimum is determined
through a complicated function of $z_0(Q)$ only. Therefore, at first
order, the low energy minimum after electroweak symmetry breaking has
to be given by
\begin{equation}
v_{\rm u,d}(Q)= v_{\rm u,d}^0 + C_{\rm u,d}\frac{ w}{m^2}~,
\end{equation}
where $C_{\rm u,d}$ are coefficients of order one and $v_{\rm u,d}^0$
are the vevs in the absence of dark energy. This is one of the main
results of this article: this explicitly determines the $Q$-dependence
of the Higgs vevs in presence of dark energy. The coefficients $C_{\rm
  u,d}$ are complicated functions of the parameters of the model such
as the gravitino mass, the gaugino mass etc \dots. Here, we do not
need their explicit expressions.

\par

As a result of the previous considerations, the masses of the atoms are
\begin{equation}
m_{_{\rm ATOM}}(Q) = m_{_{\rm ATOM}}^0 + C_{_{\rm ATOM}} \frac{w}{m^2}~,
\end{equation}
where $C_{_{\rm ATOM}}$ depends on the type of atom and is of order
one while $m_{_{\rm A}}^0$ is the atomic mass in the absence of dark
energy. Again, such an expansion is entirely due to the expansion of
$\langle z \rangle$. As a consequence, the gravitational coupling is
\begin{equation}
\alpha_{_{\rm ATOM}}= C_{_{\rm ATOM}} \frac{\partial_{\kappa_4 Q_{\rm n}} w}
{m^2 m_{_{\rm ATOM}}^0}\, .
\end{equation}
For $Q_{\rm n} \sim \mpl$, this is negligible as it behaves like
\begin{equation}
\label{eq:alphashift}
\alpha_{_{\rm ATOM}} \simeq \frac{w_{\rm now}}{m^2 m_{_{\rm ATOM}}^0}
\simeq 10^{-70} \left(\frac{m_{3/2}}{100 \, \mbox{GeV}}\right)^{-2}
\left(\frac{m_{_{\rm ATOM}}^0}{1 \, \mbox{GeV}}\right)^{-1}~.
\end{equation}
Hence, as announced, if the dark sector is shift symmetric, dark
energy decouples from baryonic matter altogether.

\par

Let us turn our attention to cold dark matter, which is composed of
neutralinos. Using the same argument as before, the mass of the
lightest supersymmetric particle can be expanded as
\begin{equation}
  m_{_{\rm CDM}}= m^0_{_{\rm CDM}}+ C_{_{\rm CDM}} \frac{w}{m^2}~,
\end{equation}
where $C_{_{\rm CDM}}$ is a dimensionless coefficient of order
one. This implies that the effective potential due to the coupling
between dark energy and dark matter can be written as
\begin{eqnarray}
V_{\rm eff}(Q) &\equiv& V_{_{\rm DE}}(Q) + n_{_{\rm CDM}} m_{_{\rm CDM}} \nonumber \\
&=& n_{_{\rm CDM}}
m_{_{\rm CDM}}^0 + \left(C_{_{\rm CDM}}  \frac{n_{_{\rm CDM}}}{m^2 \kappa_4} -2\sqrt 3
{\rm e}^{\kappa_4^2 \vert z_0\vert^2}  m^2\right) \kappa_4 w(Q)~,
\end{eqnarray}
where we have used the expression~(\ref{eq:darkenergypot}) of the dark
energy potential. The number density of dark matter particles, $n_{_{\rm
  CDM}}$, can be estimated as
\begin{equation}
n_{_{\rm CDM}}= \left(1+z \right)^3 \frac{\Omega_{_{\rm CDM}}\rho_{\rm cri}}
{m_{_{\rm CDM}}^0}~,
\end{equation}
implying that the coupling between dark matter and dark energy plays a
role for a redshift larger than
\begin{equation}
1+z \simeq  \left( \frac{m_{3/2}^2 m_{_{\rm CDM}}^0}{\mpl H_0^2 } \right)^{1/3}~,
\end{equation}
where $m\sim m_{_{\rm CDM}}^0\sim 1$ TeV leads to $z\sim 10^{20}$, \ie
the coupling between dark energy and dark matter is always negligible.

\par

Finally, we analyse the corrections to the scalar potential induced by
the coupling of the dark energy field to matter since, so far, we have
only considered the model at the classical level. Let us examine the
loop corrections to the dark energy potential induced by the MSSM
masses (bosons and fermions) which we parameterise as
\begin{equation}
m_i= m_i^0 + C_i \frac{w}{m^2}\, .
\end{equation}
where $C_i$ are species dependent constants. The Coleman--Weinberg
potential gives the one-loop correction to the scalar potential and
reads
\begin{equation}
  V_{\rm 1loop}= \frac{1}{32\pi^2} {\rm Str} \left(m^2_i\right)\Lambda_{\rm c}^2 +
  \frac{1}{64\pi^2} {\rm Str }\left[m^4_i \ln
    \left(\frac {m^2_i}{\Lambda_{\rm c}^2}\right)\right]\, ,
\end{equation}
where $\Lambda_{\rm c}$ is the cut-off of the theory and the symbol
``${\rm Str}$'' denotes the sum over all the bosons minus the sum over
all the fermions (one should not confuse $m_i$, the MSSM masses, with
$m$, the SUSY breaking mass). The main correction comes from the
quadratic divergence and reads
\begin{equation}
  \delta V_{_{\rm DE}}= \frac{1}{16\pi^2} {\rm Str }\left(m^0 _iC_i\right)
  \frac{ \Lambda^2_{\rm c}}{m^2} w\, .
\end{equation}
This term renormalises the scale appearing in the dark energy
potential, while preserving the functional form of the potential, see
Eq.~(\ref{eq:darkenergymin}). In other words, under the effect of the
radiative corrections we find that a superpotential of the form $w(Q)=
M_0^3 f(\kappa_4 Q)$, where $f$ is a dimensionless function, becomes
\begin{equation}
  w(Q)= \left[1- \frac{1}{32\sqrt 3 \pi^2}
    {\rm e}^{-\kappa_4^2 \vert z_0\vert^2} {\rm Str }\left(m^0_i C_i\right)
    \frac{ \Lambda^2_{\rm c}}{\kappa_4 m^4}\right]M_0^3 f(\kappa_4 Q)\, .
\end{equation}
The correction term can be large if the cut-off scale is of order of
the GUT scale. The main point is that the functional form of the
potential is not modified and one can absorb the radiative correction
in a redefinition of the scale $M_0$
\begin{equation}
M_0^3 \to \left[1- \frac{1}{32\sqrt 3 \pi^2}
{\rm e}^{-\kappa_4^2 \vert z_0\vert^2} {\rm Str }\left(m^0_i C_i\right)
\frac{ \Lambda^2_{\rm c}}{\kappa_4 m^4}\right]M_0^3\, .
\end{equation}
As in the usual renormalisation programme, the physical scale is the
one including the radiative corrections and not the bare one appearing
in the original Lagrangian. We do not know if this property can be
extended to all loops.

\par

To conclude this section, let us recap our main findings. We have
shown that requiring the existence of a shift symmetry in the dark
sector allows us to design a model where, at the same time, the
runaway shape of the dark energy potential is preserved and the
coupling between quintessence and the observable sector (ordinary and
dark matter) is made negligible and, hence, compatible with the local
tests of gravity. We have also shown that the shape of the potential
is not modified by the quantum corrections, at least at one loop.

\section{Breaking the Shift Symmetry}
\label{sec:breakingshift}

We have just seen that the existence of an exact shift symmetry
implies an effective decoupling between dark energy and matter.  A
problem springs from the sensitivity of the model to higher order
operators in the K\"ahler potential which break the shift
symmetry. Therefore, we reconsider the calculation of the previous
section but now relax the assumption that the K\"ahler potential is
shift symmetric. Then, the new minimum of the potential is given by
\begin{eqnarray}
\label{eq:mingene}
\kappa _4 \delta z &=& -\frac{1}{2\sqrt{3}}
\Biggl\{\sqrt{3}\kappa _4^2\left(K^{-1}_{_{\rm DE}}\right)^{Q^{\dagger }Q}
\left(\partial _QK_{_{\rm DE}}\right)^2
+\Biggl[2\sqrt{3}\left(1-\sqrt{3}\right)
\nonumber \\ & &
+\left(2\sqrt{3}-1\right)
\kappa _4^2\left(K^{-1}_{_{\rm DE}}\right)^{Q^{\dagger }Q}
\left(\partial _QK_{_{\rm DE}}\right)^2\Biggr]\frac{\kappa _4w}{m^2}
+\left(2\sqrt{3}-1\right)
\nonumber \\ & & \times
\kappa _4\left(K^{-1}_{_{\rm DE}}\right)^{Q^{\dagger }Q}
\partial _QK_{_{\rm DE}}\frac{w'}{m^2} \Biggr\}
\left[1+\kappa _4^2\left(K^{-1}_{_{\rm DE}}\right)^{Q^{\dagger }Q}
\left(\partial _QK_{_{\rm DE}}\right)^2\right]^{-1}\, .
\end{eqnarray}
Several comments are in order at this point. Firstly, in the shift
symmetric case, one has $\partial _QK_{_{\rm DE}}=0$ in the dark
energy direction and the above formula reduces to
Eq.~(\ref{eq:darkenergymin}) as expected. Secondly,
Eq.~(\ref{eq:mingene}) is in fact universal in the sense that nothing
has been assumed about the dark sector. The previous expression of the
minimum relies on an expansion in $\kappa _4w/m^2$ only. This
expansion is an extremely good approximation as we have already shown
that $\kappa _4w/m^2\simeq 10^{-88}$, thanks to the hierarchy between
the SUSY breaking scale and the dark energy scale (\ie the
cosmological constant scale). Therefore, it represents a general
expression for the position of the minimum in the Polonyi model in
presence of dark energy regardless of the precise form of the dark
sector. Thirdly, one notices in Eq.~(\ref{eq:mingene}) the presence of
a ``zeroth order term'', \ie a term which is not proportional to
$\kappa _4w/m^2$ or to $w'/m^2$, namely $\sqrt{3}\kappa
_4^2\left(K^{-1}_{_{\rm DE}}\right)^{Q^{\dagger }Q} \left(\partial
  _QK_{_{\rm DE}}\right)^2$. This means that, even if $\kappa _4w/m^2$
is extremely small, there is still a correction originating from the
dark sector K\"ahler potential. Moreover, since $\kappa_4\delta z\ll
1$, as it was perturbatively determined, one must require for
consistency that $\kappa _4^2\left(K^{-1}_{_{\rm
      DE}}\right)^{Q^{\dagger }Q} \left(\partial _QK_{_{\rm
      DE}}\right)^2\ll 1$. Otherwise, one should solve numerically the
higher order algebraic equation which controls the position of the new
minimum.

\par

Let us now determine the shape of the dark energy
potential. Straightforward calculations lead to
\begin{eqnarray}
V_{_{\rm DE}}(Q) &=& m^4{\rm e}^{\kappa _4^2\left(z_0^2+K_{_{\rm DE}}\right)}
\Biggl\{\kappa _4^2\left(K^{-1}_{_{\rm DE}}\right)^{Q^{\dagger }Q}
\left(\partial _QK_{_{\rm DE}}\right)^2+
\Biggl[2\kappa _4^2\left(K^{-1}_{_{\rm DE}}\right)^{Q^{\dagger }Q}
\nonumber \\ & & \times
\left(\partial _QK_{_{\rm DE}}\right)^2
-2\sqrt{3}\Biggr]
\frac{\kappa _4w}{m^2}
+2\kappa _4\left(K^{-1}_{_{\rm DE}}\right)^{Q^{\dagger }Q}
\partial _QK_{_{\rm DE}}\frac{w'}{m^2}\Biggr\}\, .
\end{eqnarray}
Again, this formula represents the expression of the dark energy
potential in the most general case. In the shift symmetric case, it
reduces to Eq.~(\ref{eq:darkenergypot}).

\par

Having established the above general results, let us now focus on the
breaking of shift symmetry. For this purpose, we now specialise the
K\"ahler potential and write
\begin{equation}
K_{_{\rm DE}} = -\frac{1}{2}\left(Q-Q^{\dagger}\right)^2
+\delta K_{_{\rm DE}}\left(Q,Q^{\dagger}\right)\, .
\end{equation}
In order to perturbatively break the shift symmetry, we consider
$\delta K_{_{\rm DE}}\left(Q,Q^{\dagger}\right)$ to be a small
correction in comparison to the shift symmetric zeroth order
term. Then, it is easy to show that the new minimum of the potential
is given by
\begin{equation}
\delta z\simeq \left(\sqrt{3}-1\right)\frac{w}{m^2}
-\frac{\kappa _4}{2}\left(\partial _Q\delta K_{_{\rm DE}}\right)^2\, ,
\end{equation}
where we have written $\left(K^{-1}_{_{\rm
      DE}}\right)^{Q^{\dagger}Q}=1$ as the non-shift symmetric
additional terms would give higher order corrections. This equation
should be compared to Eqs.~(\ref{eq:darkenergymin})
and~(\ref{eq:mingene}). Of course, if $K_{_{\rm DE}}$ is shift
symmetric, then the second term vanishes and one recovers
Eq.~(\ref{eq:darkenergymin}). As already noticed, in order for the
calculation to be consistent, the second term should be small $\kappa
_4^2 \left(\partial _Q\delta K_{_{\rm DE}}\right)^2\ll 1$ although not
necessarily of the same order as $\kappa _4w/m^2$. However, this is
sufficient to neglect ``second order'' terms of the form ``$\kappa
_4^2 \left(\partial _Q\delta K_{_{\rm DE}}\right)^2\times \kappa
_4w/m^2$''. Then, the potential $V_{_{\rm DE}}(Q)$ now contains only
two parts and can be expressed as:
\begin{equation}
  V_{_{\rm DE}}(Q)\simeq -2\sqrt{3}m^4 {\rm e}^{\kappa _4^2z_0^2}
  \left( \frac{\kappa_4w}{m^2} \right)
  + m_{3/2}^2
  \left(\partial_Q \delta K_{_{\rm DE}}\right)^2\, ,
\end{equation}
where $\kappa _4z_0=\sqrt{3}-1$ and we have used the fact that
$m^4\kappa_4^2\simeq m_{3/2}^2$. The first part depends only on the
superpotential $w$, and is nothing but Eq.~(\ref{eq:darkenergypot}),
whereas the second part depends only on the K\"ahler potential $\delta
K_{_{\rm DE}}$. The second term has its origin in the breaking of the
shift symmetry. In the following we will discuss the cosmological
dynamics of the $Q$--field. Let us now be slightly more specific and
write the corrections to the K\"ahler potential in the following form:
\begin{equation}
\delta K_{_{\rm DE}} = c_p \frac{(Q + Q^\dagger)^p}{\mpl^{p-2}}\, .
\end{equation}
In this case, the dimensionless parameter $\kappa _4^2 \left(\partial
  _Q\delta K_{_{\rm DE}}\right)^2\simeq c_p^2(Q/\mpl)^{2p-2}$ which
means that a choice such that $c_p\lta 10^{-2}$ is enough to guarantee
the validity of our approximation up to $\kappa_4Q \simeq {\cal
  O}(1)$. Furthermore, we assume that the first term in the potential
above has the form $M^{4+n}/Q^n$. In this case, the resulting full
potential reads
\begin{equation}
\label{eq:pottwobranches}
V_{_{\rm DE}}(Q)= \frac{M^{4+n}}{Q^n}
+ m_{3/2}^2\mpl^2c_p^2p^22^{2p-2}
\left(\frac{Q}{\mpl}\right)^{2p-2}\, .
\end{equation}
The potential possesses two branches: it goes as $Q^{-n}$ for very
small vevs and as $Q^p$ for large vevs. In between there is a minimum
and the overall runaway shape is lost. If we assume that this minimum
is located at a vev which is small in comparison to the Planck mass,
then we can approximate the denominator in the above expression by
one. Under this assumption, the scalar field value at the minimum is
\begin{equation}\label{eq:Qmin_exact}
  \left(\frac{Q_{\rm min}}{\mpl}\right)^{n+2p-2}
  \simeq \frac{M^{4+n}}{m_{3/2}^2 \mpl^{2+n}}\, .
\end{equation}
The mass scale $M$ can be fixed by assuming that, at the minimum at
the present time, $V(Q_{\rm min}) \simeq \rho_{\rm cri}$, where we
recall that $\rho_{\rm cri}$ is the critical density now. This gives
\begin{equation}\label{eq:fixM}
M^{4+n} \simeq
\left(\frac{\rho_{\rm cri}}{\mpl^2 m_{3/2}^2}\right)^{n/(2p-2)}\rho _{\rm cri}
\mpl^n\, .
\end{equation}
Using this result in Eq.~(\ref{eq:Qmin_exact}) and ignoring again
numbers of order one, we obtain
\begin{equation}\label{eq:Qmin}
\frac{Q_{\rm min}}{\mpl}\simeq \left( \frac{\rho_{\rm cri}}
{\mpl^2 m_{3/2}^2}\right)^{1/(2p-2)}
\simeq \left(\frac{H_0}{m_{3/2}}\right)^{1/(p-1)}\, .
\end{equation}
This equation implies that the value of $Q_{\rm min}$ is much smaller
than the Planck mass. This also justifies {\it a posteriori} our
assumption to approximate the denominator in
Eq.~(\ref{eq:pottwobranches}) by one.

\par

Having determined the parameters of the model in the case where the
shift symmetry is broken, let us now see whether the results of the
previous section are preserved. Firstly, we notice that the mass of
the quintessence field at the minimum is given by
\begin{equation}\label{eq:mass}
m_Q^2 \simeq m_{3/2}^{2/(p-1)}H_0^{(2p-4)/(p-1)}\,
\end{equation}
This mass is very small in comparison to $10^{-3}\mbox{eV}$ and,
therefore, may lead to a fifth force when the dark energy field sits
at the minimum. As a consequence, one needs to recompute the
gravitational coupling. Since the shift symmetry is broken, one should
now use Eq.~(\ref{eq:alphaatom}). This leads to
\begin{equation}
  \alpha_{_{\rm ATOM}} \simeq \kappa_4\partial _Q\delta K_{_{\rm DE}}\simeq
  c_p\frac{H_0}{m_{3/2}}\simeq 10^{-44}c_p\left(\frac{m_{3/2}}
{100\, \mbox{GeV}}\right)^{-1}\ll 1\, .
\end{equation}
This implies that the models are safe gravitationally and no fifth
force is present. This formula should be compared to
Eq.~(\ref{eq:alphashift}). As expected, we see that $\alpha _{_{\rm
    ATOM}}$ is larger in the non shift symmetry case but the
remarkable thing is that it is still very small. Therefore, the
gravitational tests are still satisfied even if the shift symmetry is
broken. A last remark is that, in principle, one should have computed
the quantity $\alpha_{_{\rm ATOM}}$ using the normalised field $Q_{\rm
  n}$ which does not coincide with $Q$ due to the shift symmetry
breaking term in the K\"ahler potential. However, for small $Q$ (as is
the case at the minimum, see above), the corrections are negligible
and the field can effectively be considered as canonically
normalised. Therefore, this justifies the previous calculation.

\par

In fact, the only major effect of the shift symmetry breaking is the
modification of the cosmological dynamics.  As we will see in the
following, the cosmological consequences of the setup just described
has a lot in common with the theory presented in
\cite{Brax:2006kg}. Let us discuss the evolution of the quintessence
field. Assuming that the initial field value of the quintessence field
just after reheating is much smaller than the minimum value
(\ref{eq:Qmin}), the potential is well approximated by an
inverse-power law potential.  As is well known, in this case there is
a particular attractor solution, given by
\begin{equation}
Q_{\rm att} = Q_{\rm p} \left(\frac{a}{a_{\rm p}}\right)^{3(1+w_{\rm B})/(n+2)},
\end{equation}
where it is assumed that the background is dominated by a fluid with
equation of state $w_{\rm B}$ and the scale factor $a$ is a function
of conformal time $\eta$, given by
\begin{equation}
a(\eta) = a_{\rm p} \left(\frac{\eta}{\eta_{\rm p}}\right)^{2/(1+3w_{\rm B})}.
\end{equation}
The constant $Q_{\rm p}$ is given by
\begin{equation}
  Q_{\rm p}^{-n-2} = \frac{18}{n^2a_{\rm p}^2 \eta_{\rm p}^2 M^{4+n}}
  \frac{1-w_{\rm Q}^2}{(1+3w_{\rm B})^2}\approx
\frac{\Omega_{\rm r}^{0} z_{\rm reh}^4
    m_{3/2}^{n/(p-1)}}{H_0^{n/(p-1)} \mpl^{n+2}},
\end{equation}
where we have used the expression of $M$ given by Eq.~(\ref{eq:fixM})
and chosen the time $\eta_{\rm p}$ to be the reheating time. In the
above formula, $z_{\rm reh}$ is the reheating redshift and $\Omega
_{\rm r}^0$ represents the present day radiation energy
density. Finally, $w_{\rm Q} = (-2+nw_{\rm B})/(n+2)$ is the equation
of state of the quintessence field.  This solution is valid as long as
$Q_{\rm att}\ll Q_{\rm min}$.  At a certain time, however, $Q_{\rm
  att}$ is comparable to $Q_{\rm min}$. This happens when the scale
factor reaches the value
\begin{equation}
  a_{\rm min} \simeq a_{\rm p} \left[ \Omega_{\rm r}^0 z_{\rm reh}^4
    \left(\frac{Q_{\rm min}}{\mpl}\right)^2\right]^{1/[3(1+w_{\rm B})]}.
\end{equation}
Using $\Omega_{\rm r}^0 = 10^{-5}$ and $z_{\rm reh} = 10^{28}$, this
corresponds to a redshift
\begin{equation}
  z_{\rm min} \simeq  10
\left(\frac{Q_{\rm min}}{\mpl}\right)^{-1/2}\, .
\end{equation}
For $m_{3/2} \simeq 1~\mbox{TeV}$, the last equation gives $z_{\rm
  min}\simeq 10^{13}$, so the field will sit at the minimum before big
bang nucleosynthesis. In this case, the quintessence field has no
dynamics and the model becomes effectively equivalent to a
cosmological constant.

\par

If the field is initially not on the attractor, there are two
possibilities.  The first one, called the undershoot case, is that
initially $Q_{\rm ini}>Q_{\rm att}$. In this case, the field remains
initially frozen at a value $Q = Q_{\rm ini}$ until $Q_{\rm ini} =
Q_{\rm att}$. Using the expression for $Q_{\rm att}$, one obtains the
redshift when this happens
\begin{equation}
z_{\rm u} \approx 10 \left(\frac{Q_{\rm min}}{\mpl}\right)^{n/4}
\left(\frac{Q_{\rm ini}}{\mpl}\right)^{-(n+2)/4}=10
\left(\frac{Q_{\rm min}}{\mpl}\right)^{-1/2}
\left(\frac{Q_{\rm min}}{Q_{\rm ini}}\right)^{(n+2)/4}.
\end{equation}
Since $Q_{\rm min}>Q_{\rm ini}$ one has $z_{\rm u}>z_{\rm min}$ and,
therefore, the attractor is always joined before the minimun is
reached. This means that the above analysis is still valid in an
undershoot situation.

\par

The second case is the overshoot case, in which $Q_{\rm ini}<Q_{\rm
  att}$. In this case, the field is initially dominated by kinetic
energy and evolves according to
\begin{equation}
  Q = Q_{\rm ini} + \mpl \sqrt{\frac{3\Omega_{Q_{\rm ini}}}{4\pi}}
  \left(1-\frac{a_{\rm ini}}{a}\right).
\end{equation}
As the kinetic energy is red-shifted away, at a certain point in time
the potential energy becomes comparable to the kinetic energy. The
field will then be frozen until it joins the attractor solution. The
frozen redshift can estimated to be
\begin{equation}
z_{\rm froz}\simeq 10 \left(\frac{Q_{\rm min}}{\mpl}\right)^{n/6}
\Omega _{Q_{\rm ini}}^{-(n+2)/12}\, .
\end{equation}
However, the field could also never reach this regime because it
``feels'' the presence of the minimum before being frozen. When the
field is dominated by kinetic energy, the minimum is felt at the
following time
\begin{equation}
z_{\rm kin \rightarrow min}\simeq 10^{10}\Omega _{Q_{\rm ini}}^{-1/6}\, .
\end{equation}
This time should be compared with $z_{\rm froz}$. One can show that
$z_{\rm froz}>z_{\rm kin\rightarrow min}$ if $\Omega _{Q_{\rm
    ini}}<(H_0/m_{3/2})^{2/(p-1)}$. However, the point is that these
two redshifts are large and that, in any case, the minimum is reached
very early in the history of the Universe (\ie before big bang
nucleosynthesis). When the field is approaching the minimum, the
potential is no longer of run-away form. One can show that the field
oscillates around the minimum and that the oscillations are damped by
the Hubble expansion [with a factor ${\rm e}^{-3(1-w_{\rm B})N_{\rm
    p}/2}$]. Indeed the situation is as described in
\cite{Brax:2006kg}, the only difference is the mass of the field.

\par

To summarise this section, for generic initial conditions, the field
will approach the minimum and settle there at very high redshift. The
field will oscillate around the minimum and  the model behaves
essentially like the standard $\Lambda$CDM model, since the coupling
to matter is suppressed in the theory presented so far. Indeed, very
quickly, the oscillations are damped and the field stays at the
minimum: the model is effectively equivalent to a cosmological
constant.

\section{A Different Ansatz}
\label{subsec:differentansatz}

We have studied the coupling of dark energy to the standard model
assuming that the three sectors- dark energy, supersymmetry breaking
and the MSSM- are decoupled and only interact gravitationally, \ie both
the K\"ahler potential and the superpotential are
\begin{equation}
K= K_{_{\rm DE}} + K_{\rm h}  + K_{\rm MSSM}\, , \quad
 W=W_{_{\rm DE}} + W_{\rm h}  + W_{\rm MSSM}\, .
\end{equation}
This implies that the standard model couplings become functions of the
dark energy field measured in Planck units, \ie if the gravitational
interactions were turned off, no coupling between the sectors would
exist. As a result, we have shown that the dark energy K\"ahler
potential must be almost shift symmetric. The breaking of the shift
symmetry can only occur via non-renormalisable interactions. In the
shift symmetric case, no gravitational consequences of the existence
of a nearly massless dark energy field can be detected and the
cosmological evolution of the universe is of the uncoupled
quintessence type. In the broken shift symmetric case, no
gravitational effect can be detected either and the cosmological evolution is
akin to the $\Lambda$CDM one. These results are intrinsically
dependent on the initial ansatz for the K\"ahler potential and the
superpotential. Introducing direct couplings between the dark energy
sectors and the other two sectors would increase the gravitational
effects unless the couplings were chosen to exactly cancel the
gravitationally induced interactions. This would appear as an
unnatural fine-tuning.

\par

It happens that there is a hidden assumption in the way we have
specified the model. Indeed, we have required that if we removed the
dark energy sector, the models would reduce to the usual hidden sector
symmetry breaking situation with two sectors coupled via gravitational
interactions.  If we were to dismiss this extra hypothesis, we could
envisage a more general (and more drastic) type of decoupling. Indeed
the supergravity Lagrangian is a function (forgetting the gauge
sector) of $G= \kappa_4^2 K + \ln \kappa_4^6 \vert W\vert^2$.  A
possible decoupling consists in separating~\cite{Achucarro:2008fk}
\begin{equation}
G=G_{_{\rm DE}} + G_{\rm h}  + G_{\rm MSSM}
\end{equation}
This implies that
\begin{equation}
K=K_{_{\rm DE}} + K_{\rm h}  + K_{\rm MSSM}\, ,\quad
 W=\kappa_4^6 W_{_{\rm DE}}  W_{\rm h}   W_{\rm MSSM}\, .
\end{equation}
From these expressions, one deduces that the scalar potential reads
\begin{equation}
V= {\rm e}^{\kappa_4^2 K}\left(V_{_{\rm DE}} + V_{\rm h}  + V_{\rm MSSM}-3
\kappa_4^{2}\vert W\vert^2\right)\, ,
\end{equation}
with
\begin{equation}
V_A= D_i W_A K_A^{i\bar j} \bar D_{\bar j} W_A^{\dagger}
\end{equation}
where $D_i W_A= \partial_i W_A + \kappa_4^2 \partial_i K_A W$ for $A$
running over the dark, hidden and observables sectors and fields
$\phi_i$ in each sector respectively.
Assuming  a small value of the
vacuum energy now such that  $V_{_{\rm DE,
    now}}\simeq \kappa_4^2{\vert W_{_{\rm DE}}\vert^2}$ for a dark
energy field around the Planck scale leads  to an upper bound on 
the gravitino mass
\begin{equation}
m_{3/2}\le  H_0,
\end{equation}
obtained by saying that the superpotentials in the hidden and MSSM
sectors must be less than the Planck scale cubed. This is a very low
value for the gravitino mass which violates the Fayet bound on the
gravitino mass $m_{3/2}\ge 10^{-5}$ eV~\cite{Fayet:1986zc}. Hence this
approach does not seem to be promising. We can conclude that our
original ansatz whereby the three different sectors are separated is
the simplest setting to model dark energy in supergravity. Of course,
more complex models could be built with particular couplings between
the different sectors. It would be very interesting to construct such
models explicitly when motivated by more fundamental theories such as
string theory.

\section{Conclusion}
\label{sec:conclusion}

In this paper we have found a way of solving one of the outstanding
problems associated with dark energy; namely how to incorporate it
in a supersymmetric model of particle physics. We have taken a model
of dark energy as a new sector on top of the
usual observable and supersymmetry breaking sectors of particle
physics phenomenology. We have found that the runaway shape of the
dark energy potential is determined by the superpotential of the dark
energy sector once a shift symmetry has been introduced. This
prevents the existence of long range fifth forces.
We have found a remarkable property:
the shape of the dark energy potential is not modified by radiative
corrections. The radiative corrections only modify the overall scale
of the superpotential and one can absorb this modification in a
redefinition of the bare superpotential. As a result, the runaway
property of the dark energy potential is not destroyed by one-loop
radiative corrections. If the shift symmetry is exact then the dark
energy retains its properties when incorporated into a supersymmetric
model. This is reminiscent of the solution to the hierarchy problem
which prompted the emergence of supersymmetric models whereby the
sensitivity of the Higgs mass to large scales disappears. This is due to the
cancellation of bosonic loops by fermionic loops in the perturbative expansion.
Of course, when supersymmetry is broken this cancellation is no longer
exact. Similarly, when the shift symmetry is
not exact, the presence of higher order corrections to the K\"ahler
potential induce a drastic modification of the scalar
potential. Indeed, it is not of the runaway type anymore but has a
minimum implying that the model behaves essentially like a
$\Lambda$-CDM model since very early in the history of the
universe. Hence the existence of a shift symmetry is crucial for both
the stability of dark energy to radiative corrections and the
compliance with gravity tests. As shift symmetries can be motivated in
string theory-- they already appear as a solution to the $\eta$
problem of inflation--, one may hope that dark energy models may be
constructed using stringy ingredients. In this case, a proper
understanding of the dark energy superpotential would be crucial: it
would lead to a direct comparison with observations such as the
measurement of the dark energy equation of state and its time
dependence.

\acknowledgments

This work is supported in part by STFC (CvdB and ACD). We thank the EU
Marie Curie Research \& Training network 'UniverseNet'
(MRTN-CT-2006-035863) for support.

\bibliography{references}

\end{document}